\documentclass[twocolumn,prl,aps,superscriptaddress]{revtex4-1}
\usepackage[T1]{fontenc}
\usepackage[latin9]{inputenc}

\usepackage{amsmath}
\usepackage{amssymb}
\usepackage{bbm}
\usepackage{braket}
\allowdisplaybreaks

\usepackage{graphicx}
\usepackage[colorlinks=true,citecolor=blue,linkcolor=blue]{hyperref}

\newcommand{\equref}[1]{Eq.~(\ref{#1})}

\newcommand{\figref}[1]{Fig.~\ref{#1}}

\newcommand{\refcite}[1]{Ref.~\onlinecite{#1}}
\newcommand{\refscite}[1]{Refs.~\onlinecite{#1}}

\newcommand{\tableref}[1]{Table~\ref{#1}}
\newcommand{\supl}{Supplementary Information S.}
\newcommand{\diff}{\mathrm{d}}

\newcommand{\pdagger}{{\phantom{\dagger}}}

\newcommand{\sign}{\,\text{sign}}

\renewcommand{\vec}[1]{\boldsymbol{#1}}

\newcommand{\ie}{i.e.~}

\begin{document}
\title{Selection rules for Cooper pairing in two-dimensional interfaces and sheets}
\author{Mathias S. Scheurer}
\email{Corresponding author. E-mail: mscheurer@tkm.uni-karlsruhe.de}
\affiliation{Institute for Theory of Condensed Matter, Karlsruhe Institute of Technology (KIT), 76131 Karlsruhe, Germany.}

\author{Daniel F. Agterberg}
\affiliation{Department of Physics, University of Wisconsin-Milwaukee, Milwaukee, Wisconsin 53211, USA.}

\author{Jörg Schmalian}
\affiliation{Institute for Theory of Condensed Matter, Karlsruhe Institute of Technology (KIT), 76131 Karlsruhe, Germany.}
\affiliation{Institute for Solid State Physics, Karlsruhe Institute of Technology (KIT), 76131 Karlsruhe, Germany.}

\date{\today}

\begin{abstract}
Thin sheets deposited on a substrate and interfaces of correlated materials offer a plethora of routes towards the realization of exotic phases of matter. 
In these systems, inversion symmetry is broken which strongly affects the properties of possible instabilities -- in particular in the superconducting channel. 
By combining symmetry and energetic arguments, we derive general and experimentally accessible selection rules for Cooper instabilities in noncentrosymmetric systems which yield necessary and sufficient conditions for spontaneous time-reversal-symmetry breaking at the superconducting transition and constrain the orientation of the triplet vector. 
We discuss in detail the implications for various different materials. For instance, we conclude that the pairing state in thin layers of Sr$_2$RuO$_4$ must, as opposed to its bulk superconducting state, preserve time-reversal symmetry with its triplet vector being parallel to the plane of the system. All pairing states of this system allowed by the selection rules are predicted to display topological Majorana modes at dislocations or at the edge of the system. 
Applying our results to the LaAlO$_3$/SrTiO$_3$ heterostructures, we find that while the condensates of the (001) and (110) oriented interfaces must be time-reversal symmetric, spontaneous time-reversal-symmetry breaking can only occur for the less studied (111) interface.
We also discuss the consequences for thin layers of URu$_2$Si$_2$ and UPt$_3$ as well as for single-layer FeSe.
On a more general level, our considerations might serve as a design principle in the search for time-reversal-symmetry-breaking superconductivity in the absence of external magnetic fields. 
\end{abstract}
\maketitle

\section{Introduction}
The realization and characterization of two-dimensional (2D) superconducting phases in various different systems constitutes a topic of great current interest \cite{Reyren,Iwasa1,FeSeObs,Iwasa2,110SC}. This is motivated by the promising role played by 2D superconductors in the search for topological Majorana modes and related applications \cite{Bernevig}, by the gate tunability of the electronic properties \cite{CavigliaDome,Iwasa1,Iwasa2}, and by the fundamental interest in superconducting transitions in reduced dimensions. 
Particularly interesting examples are given by LaAlO$_3$/SrTiO$_3$ heterostructures, that show very rich electronic behavior \cite{MannhartReview}, and single-layer FeSe on [001] SrTiO$_3$ with significantly enhanced transition temperatures compared to its bulk value \cite{FeSeObs}.
This also motivates closer inspection of superconducting thin films of other correlated materials such as Sr$_2$RuO$_4$ \cite{Sr2RuO4ThinFilms} and UPt$_3$ \cite{UPt3ThinFilms} which, in addition, promises to offer insights into the electronic properties of the bulk material.

\begin{figure}[t]
\begin{center}
\includegraphics[width=\linewidth]{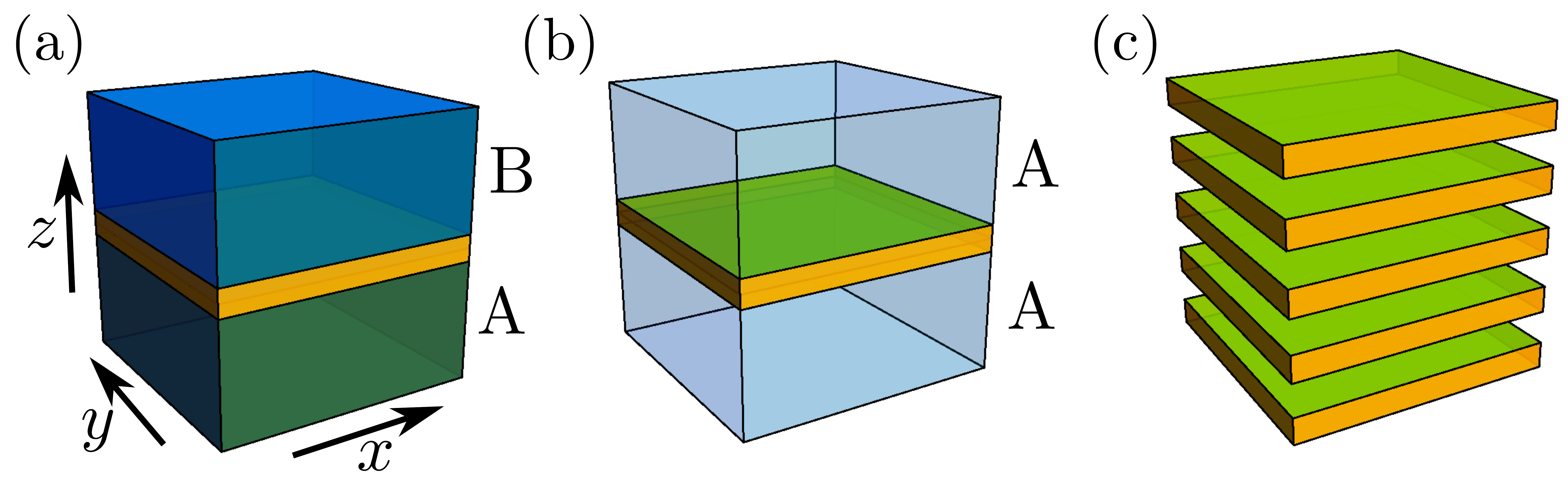}
\caption{From a symmetry point of view, 2D systems (yellow) can be grouped into those realized in an asymmetric (a) or symmetric (b) environment. Layered materials consisting of weakly coupled sheets as shown in (c) are not included in our definition of 2D systems.}
\label{DifferentSetups}
\end{center}
\end{figure}

Noncentrosymmetric 2D superconductors form a particularly important class since inversion symmetry is naturally broken in the practical realization of 2D systems: As shown in \figref{DifferentSetups}(a), both for an interface as well as for a thin layer on a substrate ($B$ is vacuum) or in an asymmetric environment ($B$ not vacuum, but $A \neq B$), inversion symmetry is broken; 
it can only be restored in the case of a thin layer in a symmetric environment as shown in \figref{DifferentSetups}(b).

A pivotal property of superconducting states is their behavior under the inversion of the time direction. Not only does it determine the topological classification \cite{SymClasses} but also essentially influences the electromagnetic and thermal response of these systems \cite{LudwigTFT}.

In this work, constraints on possible pairing states of noncentrosymmetric systems are derived that follow from the combination of symmetry and energetic arguments. These ``symergetic'' selection rules state that any superconducting order parameter transforming under a complex or multidimensional representation of the point group $\mathcal{G}$ of the normal state necessarily breaks time-reversal symmetry (TRS). For 2D systems, it furthermore holds that this is only possible if $\mathcal{G}$ contains a threefold rotation symmetry. Finally, if $\mathcal{G}$ includes a twofold rotation symmetry $C_2^\perp$ perpendicular to the plane of the system, the component of the triplet vector along the axis of $C_2^\perp$ must vanish.
These results hold under the assumption that \emph{(i)} the energetic splitting $E_{\text{so}}$ of the Fermi surfaces is larger than the superconducting order parameter and \emph{(ii)} that the superconducting
phase does not break translation invariance. 
We emphasize that our notion of 2D does not include strongly anisotropic three-dimensional systems illustrated in \figref{DifferentSetups}(c).
  
In the following, we will first present a proof of the selection rules stated above and then discuss the consequences for the time-reversal and topological properties of possible pairing states in several different 2D materials.  

\section{Results}
In order to decide which superconducting states are possible, we consider a system with pairing Hamiltonian 
\begin{equation}
H_{\text{MF}}=\sum_{\vec{k}}\left[\psi_{\vec{k}}^{\dagger}h_{\vec{k}}\psi^{\phantom{\dagger}}_{\vec{k}}+\frac{1}{2}\left(\psi_{\vec{k}}^{\dagger}\Delta_{\vec{k}}\left(\psi_{-\vec{k}}^{\dagger}\right)^{T}+ \,\text{H.c.}\right)\right], \label{GeneralMFHamiltonian}
\end{equation}
already taking into account assumption \emph{(ii)}. The fermionic creation and annihilation operators $\psi_{\vec{k}}^{\dagger}$
and $\psi_{\vec{k}}$ are $N$-component spinors that describe the spin and orbital degrees of freedom as well as potentially relevant subbands that result from the confinement along the direction perpendicular to the plane of the system. Correspondingly, the normal state Hamiltonian $h_{\vec{k}}$ and the pairing function $\Delta_{\vec{k}}$ in \equref{GeneralMFHamiltonian} are $N\times N$ matrices.
Note that this approach and the following analysis goes beyond the pseudospin description that is commonly used \cite{SigristUeda,AgterbergSigrist,Samokhin2} for studying pairing in systems with spin-orbit coupling.

To begin with the constraints resulting from \textit{symmetries}, let us investigate the transformation properties of $\Delta_{\vec{k}}$ under time-reversal and the elements $g$ of
the point group $\mathcal{G}$ of the normal state. The time-reversal operator is given by $\Theta=T\mathcal{K}$ with unitary $T$ and $\mathcal{K}$ denoting complex conjugation. Time-reversal acts on the pairing field according to
\begin{equation}
\Delta_{\vec{k}}\overset{\Theta}{\longrightarrow}T\Delta_{-\vec{k}}^{*}T^T.\label{eq:TR}
\end{equation}
Under a point group operation $g\in\mathcal{G}$ holds $\Delta_{\vec{k}}\overset{g}{\longrightarrow}R_{\psi}(g)\Delta_{R^{-1}_{v}(g)\vec{k}}R_{\psi}^{T}(g)$, where $R_{v}(g)$ and $R_{\psi}(g)$ transform vectors and spinors, respectively.

To identify the order parameter, we expand the pairing field 
\begin{equation}
\Delta_{\vec{k}}=\sum_{n}\sum_{\mu=1}^{d_{n}}\eta_{\mu}^{n}\chi_{\vec{k}\mu}^{n}\label{eq:expans}
\end{equation}
with respect to the basis of $N\times N$ matrix fields $\chi_{\vec{k}\mu}^{n}$
transforming under the irreducible representation (IR) $n$ of $\mathcal{G}$. Here $d_{n}$ is the dimensionality of the IR $n$ and $\eta_{\mu}^{n}$ are complex-valued coefficients. 
Note that, before analyzing fluctuations, we first have to determine the form and, in particular, the symmetry properties, of possible order parameters which is the central theme of the present work. Including fluctuations will modify the behavior of physical quantities in the vicinity of the phase transition. 
As will be seen below, the superconducting transitions we investigate are always second order on the mean-field level and, hence, a Ginzburg-Landau (GL) expansion can be used to determine the candidate pairing states.
Taking into account the usual orthogonality relations of IRs \cite{Lax}, the free energy $\mathcal{F}$ assumes the form
\begin{equation}
\mathcal{F}\hspace{-0.2em}\left[\eta_{\mu}^{n}\right]=\mathcal{F}\hspace{-0.1em}\left[0\right]+\sum_{n}\sum_{\mu=1}^{d_{n}}a_{n}(T)\left|\eta_{\mu}^{n}\right|^{2} + \mathcal{F}_{\geq 4}\hspace{-0.2em}\left[\eta_{\mu}^{n}\right] \label{GLExpansion}
\end{equation}
with higher order terms $\mathcal{F}_{\geq 4} \in \mathcal{O}\left(|\eta|^4\right)$.
The coefficient $a_{n_{0}}(T)$ that first changes sign
determines the IR $n=n_{0}$ of the order parameter.
If the representation $n_{0}$ is real, the TRS of the normal state implies that the matrix fields $\chi_{\vec{k}\mu}^{n} T^\dagger$ can be chosen to be Hermitian (see \supl 1) and from \equref{eq:TR} follows $\eta_{\mu}^{n}\overset{\Theta}{\longrightarrow} \pm (\eta_{\mu}^{n})^*$ for $\Theta^2 = \mp \mathbbm{1}$.
As the global phase of the order parameter can always be absorbed by a $U(1)$ transformation of the fields, we need at least a two-dimensional ($d_{n_0}>1$) order parameter vector with a nontrivial relative phase to break TRS. As can be seen, e.g., in Table~\ref{PossibleSCsSrRuO}, where all symmetry-allowed order parameters for the point group $C_{4v}$ \cite{Lax} are summarized, only the pairing state $e_{(1,i)}$ transforming under the 2D IR $E$ breaks TRS.
Note that this is different in the case of complex representations, where time-reversed partners transform according to different IRs. Consequently, we have to identify pairing states either in complex or in multi-dimensional IRs to obtain a TRS-breaking superconductor.

\renewcommand{\arraystretch}{1.3}
\begin{table}[tb]
\begin{center}
\caption{Possible pairing states for a system with $C_{4v}$ point group. It is indicated whether the phase preserves (y) or breaks (n) TRS. Here $X$ and $Y$ are continuous functions on the whole Brillouin zone transforming as the in-plane momenta $k_x$ and $k_y$. For future reference, we also show the associated triplet component of the order parameter with $\sigma_j$ denoting Pauli matrices in spin space. Although our analysis is more general, we here focus, for simplicity, on pairing states that transform trivially in orbital/subband space.}
\label{PossibleSCsSrRuO}
 \begin{tabular} {cc|c|c|c} \hline \hline
   Gr.~th.    & Pairing & Symmetry & TRS  & $\vec{d}_{\vec{k}}\cdot\vec{\sigma}$  \\ \hline
 $A_1$ & $s$-wave &   $1, X^2+Y^2$  & y   & $Y\sigma_x-X\sigma_y$             \\ 
$A_2$ & $g$-wave &   $XY(X^2-Y^2)$  & y   & $X\sigma_x+Y\sigma_y$             \\ 
$B_1$ & $d_{x^2-y^2}$ &   $X^2-Y^2$  & y   & $Y\sigma_x+X\sigma_y$            \\ 
$B_2$ & $d_{xy}$  &   $XY$  & y     & $X\sigma_x - Y\sigma_y$       \\  \hline
$E(1,0)$ & $e_{(1,0)}$ &   $X$  & y   & $\sigma_z Y$              \\ 
$E(1,1)$ & $e_{(1,1)}$  &   $X+Y$ & y   &  $\sigma_z (Y-X)$          \\ 
$E(1,i)$ & $e_{(1,i)}$ &   $X+iY$   & n  & $\sigma_z (X+iY)$     \\ \hline \hline
 \end{tabular}
\end{center}
\end{table}
\renewcommand{\arraystretch}{1}

\subsection{Weak-pairing limit}
To deduce the consequences resulting from the \textit{energetic} assumption \emph{(i)}, it is convenient to diagonalize the free Hamiltonian $h_{\vec{k}}$ 
by the unitary transformation $\psi_{\vec{k}i}=\sum_{a}\left(\phi_{\vec{k}a}\right)_{i}f_{\vec{k}a}$
that is made of its eigenfunctions $\phi_{\vec{k}a}$ satisfying $h_{\vec{k}}\phi_{\vec{k}a}=\varepsilon_{\vec{k}a}\phi_{\vec{k}a}$.
Since $h_{\vec{k}}$ is time-reversal symmetric, \ie $\Theta h_{\vec{k}}\Theta^{-1}=h_{-\vec{k}}$, we know that $\Theta\phi_{\vec{k}a}$
is an eigenstate of $h_{-\vec{k}}$ with the same energy. The broken
inversion symmetry at the interface along with spin-orbit coupling
further imply that the Fermi surfaces are non-degenerate in the generic case.
This implies for the wave functions that \cite{CrunoePhaseFactor}
\begin{equation}
\phi_{\vec{k}a}=e^{i\varphi_{\vec{k}}^{a}}\Theta\phi_{-\vec{k}a},\label{eq:phasefacts}
\end{equation}
where the phase factors must satisfy the condition $e^{i\varphi_{\vec{k}}^{a}}= \mp e^{i\varphi_{-\vec{k}}^{a}}$
as a consequence of $\Theta^{2}=\mp \mathbbm{1}$. 

The Hamiltonian can now be cast in the quadratic form $H=\frac{1}{2}\sum_{\vec{k}}\Psi_{\vec{k}}^{\dagger}h_{\vec{k}}^{\text{BdG}}\Psi_{\vec{k}}$ using the Nambu
spinor $\Psi_{\vec{k}a}=(f^{\phantom{\dagger}}_{\vec{k}a},f_{-\vec{k}a}^{\dagger}e^{-i\varphi_{\vec{k}}^{a}})^{T}$.
The off-diagonal elements of the associated Bogoliubov-de Gennes (BdG) Hamiltonian, characterizing the superconducting state,
are given by $D_{\vec{k}ab}=\braket{\phi_{\vec{k}a}|\Delta_{\vec{k}}T^{\dagger}|\phi_{\vec{k}b}}$.

\begin{figure}[tb]
\begin{center}
\includegraphics[width=0.6\linewidth]{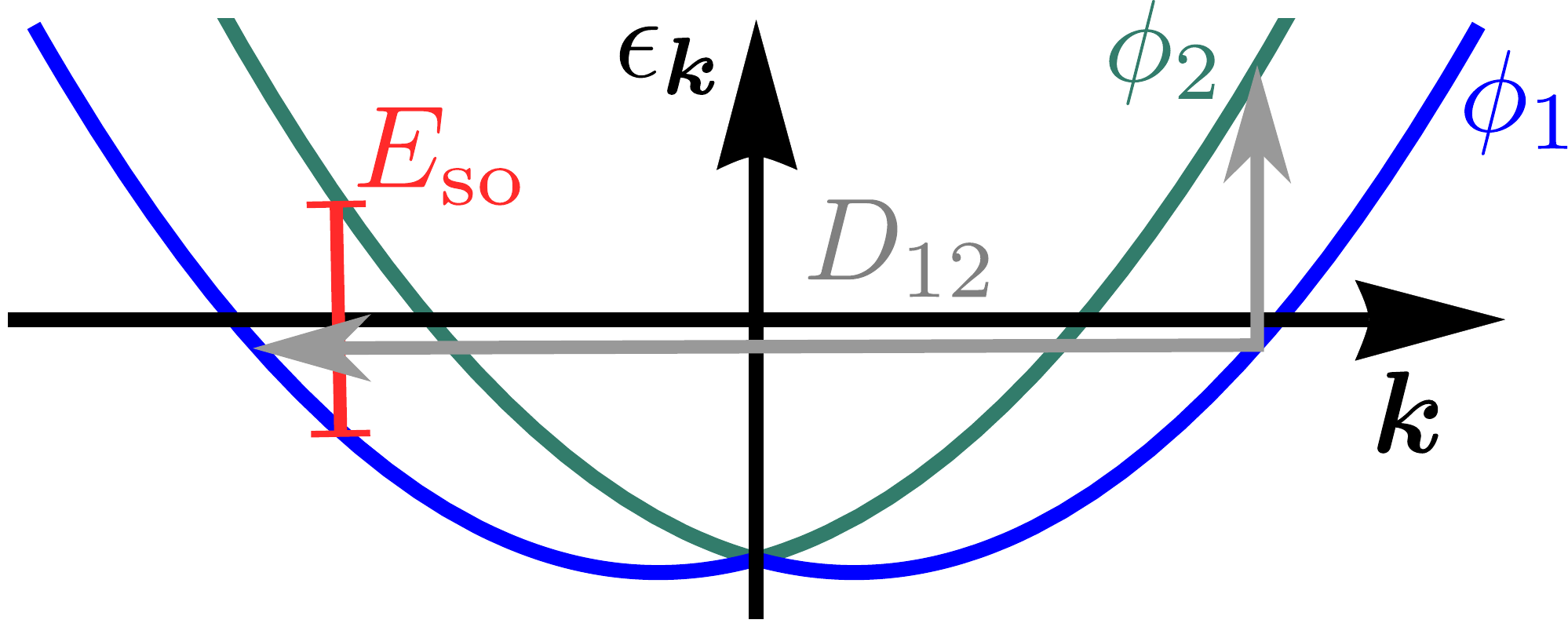}
\caption{Due to assumption (\emph{i}), the inter-Fermi surface matrix elements (gray arrow) can be neglected as they couple states separated by energies of order $E_{\text{so}}$.}
\label{WeakPairingApproximation}
\end{center}
\end{figure}

We now consider the \emph{weak-pairing limit} (see \figref{WeakPairingApproximation})
that implies that partners of a Cooper pair always originate within
a given Fermi sheet and not between states of different sheets. If
this is the case, it holds 
\begin{equation}
 D_{\vec{k}ab}=\widetilde{\Delta}_{\vec{k}a}\delta_{ab}.
\end{equation} 
Note, this assumption does not exclude frequently discussed pairing
states that are due to interband interactions. It merely requires
that anomalous averages are made of the same quantum numbers as the
normal state. 
In this weak-pairing limit, we immediately obtain the eigenvalues
of the BdG Hamiltonian $h_{\vec{k}}^{\text{BdG}}$ as $E_{\vec{k}a}=\pm (\varepsilon_{\vec{k}a}^2+|\widetilde{\Delta}_{\vec{k}a}|^{2})^{1/2}$,
i.e.~$|\widetilde{\Delta}_{\vec{k}a}|$ is the superconducting gap on
the Fermi surface. 

The behavior of $\widetilde{\Delta}_{\vec{k}a}$ under point group operations follows
from inserting \equref{eq:expans} and using that the wave functions
of non-degenerate Fermi surfaces must transform as $\phi_{\vec{k}a}=e^{i\rho_{\vec{k}}^{a}}R_{\psi}^{\dagger}(g)\phi_{R_{v}(g)\vec{k}a}$
with phase factors $e^{i\rho_{\vec{k}}^{a}}$. We obtain that the basis
functions $\varphi_{\vec{k}a}^{\mu n} := \braket{\phi_{\vec{k}a}|\chi_{\vec{k}\mu}^{n}T^{\dagger}|\phi_{\vec{k}a}}$
transform under the same, $a$-independent, IRs
as the matrix fields $\chi_{\vec{k}\mu}^{n}$. Thus, once we have
found the IR $n_{0}$ under which the pairing
field $\Delta_{\vec{k}}$ transforms, along with the associated
order parameter vector $\vec{\eta}^{n_0}=\left(\eta_{1}^{n_{0}},\cdots,\eta_{d_{n_{0}}}^{n_{0}}\right)$,
we also know the symmetry properties of the gap function 
\begin{equation}
\widetilde{\Delta}_{\vec{k}a}=\sum_{\mu=1}^{d_{n_{0}}}\eta_{\mu}^{n_{0}}\varphi_{\vec{k}a}^{\mu n_{0}},
\end{equation}
as it transforms exactly the same way.

\subsection{Microscopic derivation of the GL expansion}
Let us first focus on a single IR $n_0$ with the associated interaction ($g>0$)
\begin{align}
 H_{\text{int}} = - g \sum_{\mu,\vec{k},\vec{k}'}  \left[ \psi^\dagger_{\vec{k}} \chi_{\vec{k}\mu}^{n_0}  \left(\psi^\dagger_{-\vec{k}}\right)^T \right] \left[ \psi^T_{-\vec{k}'} \left(\chi_{\vec{k}'\mu}^{n_0}\right)^\dagger \psi^\pdagger_{\vec{k}'} \right] \label{InteractinHamInNo}
\end{align} 
in the Cooper channel. One can write down the GL expansion in the weak-pairing limit to all orders in the order parameter $\widetilde{\Delta}_{\vec{k}a}$ formally expressed in terms of Fermi surface averages $\braket{|\widetilde{\Delta}_{\vec{k}a}|^{2l}}_a$.
As discussed in more detail in the \supl 2, resummation shows that $\mathcal{F}_{\geq 4}\hspace{-0.2em}\left[\eta_{\mu}^{n}\right] \geq 0$ and, hence, the superconducting transition must be second order on the mean-field level as long as the normal phase is time-reversal symmetric.
This justifies focusing on the first few orders of the GL expansion to deduce constraints on possible pairing states. 
To fourth order, it holds $\mathcal{F}_{\geq 4}\hspace{-0.2em}= I |\vec{\eta}^{n_0}|^4 \beta(\vec{z}^{n_0}) + \dots$, where $\vec{z}^{n_0}:=\vec{\eta}^{n_0}/|\vec{\eta}^{n_0}|$, $I > 0$ is a (temperature-dependent) prefactor, and
\begin{equation}
 \beta(\vec{z}^{n_0}) = \sum_a \left\langle \Bigl|\sum_{\mu =1}^{d_{n_0}} z^{n_0}_\mu \varphi_{\vec{k}a}^{\mu n_0}\Bigr|^{4} \right\rangle_{\hspace{-0.3em}a\hspace{0.2em}} \label{BetaCoefficient}
\end{equation} 
has been introduced.
Let us assume that the minimum occurs at $\vec{z}^{n_0} \in \mathbbm{R}^{d_{n_0}}$ and define $\vec{z}^{n_0}_{\xi} := \vec{z}^{n_0}|_{\vec{z}^{n_0}_{\mu_0} \rightarrow \vec{z}^{n_0}_{\mu_0} e^{i\xi}} $ for some $\mu_0$ with $\vec{z}^{n_0}_{\mu_0} \neq 0$. It follows from \equref{BetaCoefficient} that
\begin{equation}
   \beta(\vec{z}^{n_0}_{\xi}) \sim \beta(\vec{z}^{n_0}) - C(\vec{z}^{n_0}) \xi^2 \label{AsymptoticForm}
\end{equation} 
as $\xi \rightarrow 0$. For $n_0$ being a real and multidimensional IR, one finds $C(\vec{z}^{n_0})>0$ (see Methods) except for $\vec{z}^{n_0} = \vec{e}_{\mu_0}$, with $\vec{e}_{\mu}$ denoting the unit vector along the $\mu$ direction, where $C=0$ following from gauge invariance. In the latter case, one has to take instead $\vec{z}^{n_0}_{\xi} \propto (\vec{e}_{\mu_0} + i \xi \vec{e}_{\mu_1})$ with $\mu_1\neq \mu_0$ again yielding $\beta(\vec{z}^{n_0}_{\xi}) < \beta(\vec{z}^{n_0})$ for small, but finite $\xi$. 
This means that introducing relative complexity between the components lowers the free energy and, hence, the order parameter must break TRS. This means that the two pairing states $e_{(1,0)}$ and $e_{(1,1)}$ in Table~\ref{PossibleSCsSrRuO} are allowed by symmetry but suppressed energetically in the weak-pairing limit. 
The analogous discussion for complex IRs, which are best though of as real reducible representations of dimension $2d_{n_0}$, can be found in the \supl 3. It yields that the superconducting state automatically breaks TRS when $n_0$ is complex (also for $d_{n_0}=1$). 

Two remarks are in order.
First, we emphasize that, although the main focus of this paper is on 2D systems, this result also holds for the three-dimensional (3D) case.
Second, it provides a tool to identify TRS-breaking noncentrosymmetric superconductors experimentally: The observation of a splitting of the superconducting transition under the influence of an external symmetry-breaking perturbation such as uniaxial strain indicates that the representation of the order parameter must be multidimensional or complex implying broken TRS.

\subsection{Further consequences for 2D systems}
Shifting $\vec{k}\rightarrow-\vec{k}$ in the weak-coupling limit of the pairing term 
and using the behavior of the phase factors in \equref{eq:phasefacts}
under this shift, we obtain the important property 
\begin{equation}
\widetilde{\Delta}_{\vec{k}a}=\pm \widetilde{\Delta}_{-\vec{k}a}\,\,\,\,{\rm if}\,\,\,\,\Theta^{2}=\mp \mathbbm{1}. \label{SymOfd}
\end{equation}
Naturally, the upper sign is most relevant for fermionic pairing, yet we include the more general behavior for two reasons: Firstly, it illustrates the importance
of normal state TRS for the fact that the gap function $\widetilde{\Delta}_{\vec{k}a}=  \braket{\phi_{\vec{k}a}|\Delta_{\vec{k}}T^{\dagger}|\phi_{\vec{k}a}}$
has a well-defined parity. Secondly, there are situations \footnote{A simple example is provided by a 2D electron gas with a strong in-plane magnetic field in the absence of spin-orbit coupling.} where fermionic TRS is broken, however, the effective low-energy theory of the system has an emergent TRS that satisfies $\Theta^2 = \mathbbm{1}$.  

Let us first focus on the upper signs in \equref{SymOfd}.
Suppose that the point group contains a twofold rotation $C_2^\perp$ with
$R_{v}=-\mathbbm{1}$, i.e., perpendicular to the plane of the system. This is only allowed in even space dimensions,
since $\det R_{v}=1$, which is why we will be focusing on 2D systems in the following. As dictated by \equref{SymOfd}, $\widetilde{\Delta}_{\vec{k}a}$ has to be an even function of $\vec{k}$ and, hence, no solutions with finite gap can occur that
are odd under this rotation. 

Before discussing in the next subsection under which conditions pairing states with vanishing intraband matrix elements are energetically disfavored, let us directly deduce the physical consequences of this selection rule. In the simplest case of just a single relevant orbital, the triplet vector $\vec{d}$ transforms as a vector under rotation forcing its component along the axis of $C_2^\perp$ to vanish in any pairing state that is even under $C_2^\perp$. 
For the case of several orbitals/subbands, where the analysis is more involved (see \supl 4), the orbital/subband-diagonal matrix elements of this component can be shown to vanish as well.


Secondly, in order to discuss TRS, we start by considering again the point group $C_{4v}$ as an example. The order parameter cannot transform under the 2D IR $E$ that is
required for TRS-breaking since $E$ is odd under $C_2^\perp$. Thus, we can exclude a finite order parameter that transforms as $k_{x}+ik_{y}$
or any other superpositions of $k_x$ and $k_{y}$ for that matter. In the case of 2D systems, the
matrix elements of the pairing field on a non-degenerate Fermi surface
are too restrictive to allow for any of these pairing states. 

It is straightforward to generalize this analysis to all possible point groups of noncentrosymmetric 2D electron systems: For analogous reasons to $C_{4v}$, TRS-breaking superconductivity is not possible for the interface point group $C_4$. The same holds for the isomorphic groups $D_4$, $D_{2d}$ and $S_4$ that describe possible symmetries of 2D electronic sheets. For all other symmetry groups without any rotation symmetry (such as for $C_{1}$, \ie in the absence of any symmetries) or containing only a twofold rotation normal to the plane, all IRs are real and one-dimensional such that TRS-breaking superconductivity is forbidden as well.
For the remaining possible noncentrosymmetric point groups of 2D electron systems, all of which contain a threefold rotation, one cannot exclude TRS-breaking superconductivity without further assumptions.

Similar reasoning implies that in the case of spinless fermions, i.e., for the lower signs in \equref{SymOfd}, the emergent TRS must be necessarily broken at the superconducting transition if the point group \footnote{This statement also holds for centrosymmetric point groups as long as the weak-pairing description applies.} contains a proper or improper fourfold rotation symmetry.

\begin{figure*}[tb]
\begin{center}
\includegraphics[width=0.7\linewidth]{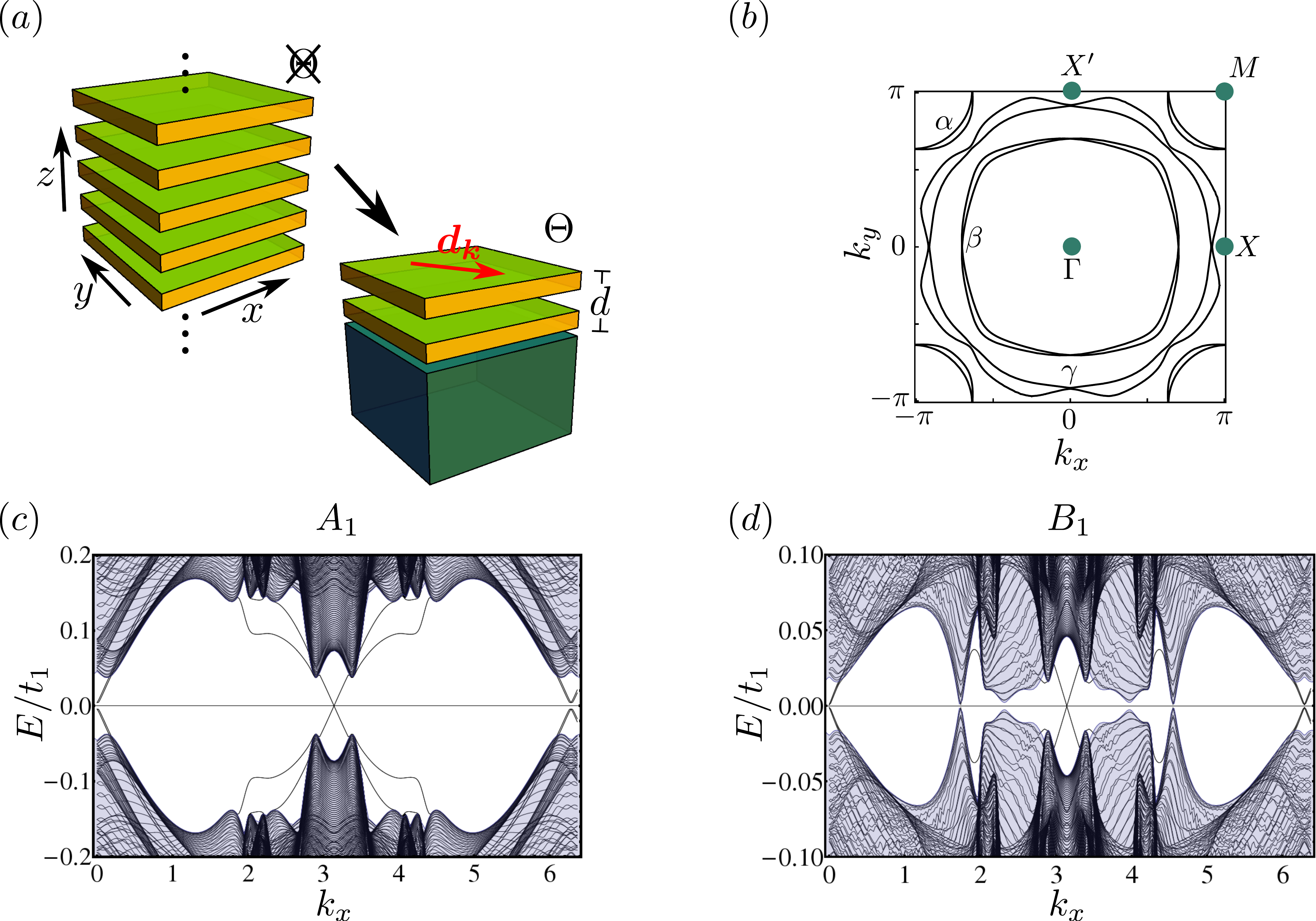}
\caption{(a) Illustration of the gedankenexperiment explained in the main text with superconductor and substrate shown in yellow and green, respectively. (b) Fermi surfaces of thin layers of Sr$_2$RuO$_4$ following from the model defined in the Methods. (c) and (d) show the low-energy part of the spectrum of the superconducting states transforming under $A_1$ and $B_1$ for open (periodic) boundary conditions along the $y$ ($x$) direction. For concreteness, we have taken pure triplet pairing with $\vec{d} \propto 0.2t_1(\sin k_2,-\sin k_1,0)^T$ and $\vec{d} \propto 0.08t_1(\sin k_2,\sin k_1,0)^T$, respectively (cf.~\tableref{PossibleSCsSrRuO}). As argued in the main text, the spectra of the $B_2$ and $A_2$ pairing states look qualitatively similar to (c) and (d).}
\label{Sr2RuO4Spectra}
\end{center}
\end{figure*}

\subsection{Beyond weak pairing}
Let us next go beyond the weak-pairing limit and clarify under which conditions a (translation-invariant) superconducting phase with a vanishing intra-Fermi-surface order parameter can occur. 

To this end, we consider the most favorable scenario where, at low energies, the effective electron-electron interaction is dominated by the Cooper channel in \equref{InteractinHamInNo} with $n_0$ being odd under the twofold rotation $C_2^\perp \in \mathcal{G}$. All other competing channels are assumed to be negligible. From the analysis above, we know that the associated matrix elements $D_{\vec{k}ab}$ vanish for $a=b$. 
By deriving a general upper bound for the energy gain $\Delta E$ when entering the superconducting phase at zero temperature,
we have shown that superconductivity with $D_{\vec{k}aa} = 0$ is only possible when the spin-orbit splitting $E_{\text{so}}$ on the Fermi surface satisfies 
\begin{equation}
 E_{\text{so}} < \frac{2\Lambda}{\sinh(1/\lambda)} \label{ConditionForWPL}
\end{equation} 
where $\lambda$ denotes the associated dimensionless coupling constant and $\Lambda$ the cutoff of the theory (see Methods for more details). 
This means that, in the weak-coupling limit, $\lambda \ll 1$, superconductivity can only emerge when the spin-orbit coupling is exponentially small. 
The physical reason is that the superconducting order parameter only couples states at energies differing by $E_{\text{so}}$ as illustrated in \figref{WeakPairingApproximation}. 
We have furthermore found that the value of the order parameter maximizing the energetic gain $\Delta E$ is larger than $E_{\text{so}}/2$ which shows the relation between the validity of the weak-pairing description and assumption \emph{(i)}.

\subsection{Consequences for Sr$_2$RuO$_4$}
The general results presented above naturally lead to the gedankenexperiment illustrated in \figref{Sr2RuO4Spectra}(a): Imagine putting a 3D bulk superconductor with a TRS-breaking order parameter on a substrate and gradually reducing its thickness $d$. If the resulting, necessarily noncentrosymmetric, point group $\mathcal{G}$ of the thin layer system does not contain a threefold rotation symmetry, either superconductivity disappears or TRS must be restored below a critical value of $d$. Furthermore, if $C_2^{\perp}\in \mathcal{G}$, the triplet vector must be aligned parallel to the plane of the system. We note that these expectations are consistent with the explicit single-band-model calculations in \refscite{FujimotoSr2RuO4,YanaseSr2RuO4}.

A natural candidate material for this gedankenexperiment is provided by Sr$_2$RuO$_4$ hosting a superconducting phase \cite{FirstSCObs} which is widely believed to be a TRS-breaking chiral $p$-wave state with triplet vector $\vec{d}_{\vec{k}} \propto (k_x+ik_y) \vec{e}_z$ \cite{KnightShiftSRO,JosephsonDiffractionSRO,KapitulnikTRSBSRO}. Its small superconducting transition temperature and the near degeneracy \cite{SigristRiceTriplet} of the chiral $p$-wave order parameter with the triplet states transforming under the 1D IRs of its bulk point group $D_{4h}$ along with the strong spin-orbit coupling of Ru make it an ideal system to apply the weak-pairing description and the selection rules derived above. Furthermore, thin layers of this material have been fabricated and shown to be superconducting \cite{Sr2RuO4ThinFilms}. As the point group is reduced to $C_{4v}$ by the presence of the substrate, we conclude that both TRS must be restored and the triplet vector must rotate to be aligned in the plane of the system upon reducing the thickness $d$.

Consequently, the superconducting condensate belongs to symmetry class DIII \cite{SymClasses} and is, hence, characterized by a $\mathbbm{Z}_2$ topological invariant $\nu$ with $\nu=-1$ ($\nu=+1$) characterizing the topological (trivial) phase \cite{Bernevig}. In the weak-pairing limit, it assumes the simple form \cite{ZhangTopInv}
\begin{equation}
 \nu = \prod_a \left(\sign(\widetilde{\Delta}_{\vec{k}_aa}) \right)^{m_a}, \label{DefinitionTopInvariant}
\end{equation} 
where the product involves all Fermi surfaces, $\vec{k}_a$ is an arbitrary momentum on and $m_a$ the number of time-reversal invariant momenta (green dots in \figref{Sr2RuO4Spectra}(b)) enclosed by Fermi surface $a$. In order to calculate the topological properties of the remaining candidate pairing states transforming under the four 1D IRs of $C_{4v}$ (see \tableref{PossibleSCsSrRuO}), we take the tight-binding model for the Ru $t_{2g}$ states with atomic spin-orbit coupling that is commonly used \cite{Scaffidi} for bulk Sr$_2$RuO$_4$ (see Methods) adding the inversion asymmetric hopping term $\delta (L_x \sin(k_y) - L_y \sin(k_x))$ allowed by the residual $C_{4v}$ symmetry. Here $L_j$ denote angular momentum operators projected onto the $t_{2g}$ manifold. The prefactor $\delta$ is nonuniversal and unknown, however, the following discussion is independent of its value as long as all $m_a$ are the same as in the limit $\delta = 0$ which holds as long as $\delta < 0.47t_1$ with $t_1$ denoting the largest centrosymmetric hopping parameter.
The resulting Fermi surfaces for the rather large value $\delta = 0.45t_1$ are shown in \figref{Sr2RuO4Spectra}(b).

In addition, we assume that the triplet component is larger than the admixed singlet component since triplet is dominant in bulk Sr$_2$RuO$_4$. 
Consequently, the sign of $\widetilde{\Delta}_{\vec{k}_aa}$ is opposite on Fermi surfaces split by the inversion asymmetric hopping $\delta \neq 0$. 
From \equref{DefinitionTopInvariant} and \figref{Sr2RuO4Spectra}(b) it then directly follows that the state transforming under $A_1$, which is fully gapped in the weak-pairing limit, is topological. This is confirmed by the spectrum shown in \figref{Sr2RuO4Spectra}(c) characterized by a gapless Kramers pair of Majorana modes crossing the bulk gap in the vicinity of $k_x = \pi$. The subgap states around $k_x = 0$ result from the $\beta$ and $\gamma$ band being topological separately. The band mixing, however, introduces a mass to the associated edge modes.
Since \equref{DefinitionTopInvariant} also holds for the 1D DIII $\mathbbm{Z}_2$ invariant (with $m_a = 1$) \cite{ZhangTopInv}, we directly find the nontrivial invariant $\nu_x = -1$ ($\nu_y = -1$) for the fictitious 1D system at $k_x=\pi$ ($k_y=\pi$). Therefore, the system is characterized by the weak indices $(1,1)$ such that a Kramers doublet of isolated Majorana modes emerges at any dislocation with Burgers vector $\vec{b} = (b_1,b_2)$ satisfying $b_1 + b_2$ odd \cite{TeoKane,VishwDisl}.

The orbital polarization and symmetry restrictions render the Fermi-surface splitting very small along the high-symmetry lines $k_x = 0,\pi$ and $k_y = 0,\pi$ as can be seen in \figref{Sr2RuO4Spectra}(b). Although being nodal in the strict weak-pairing limit, the $B_2$ pairing state is thus fully gapped for values of the order parameter that are larger than the small splitting at these high-symmetry lines but much smaller than the average value of $E_{\text{so}}$ such that the symergetic arguments presented above still apply. For this reason, the $B_2$ state has the same topological signatures as the $A_1$ order parameter.

However, the $B_1$ and $A_2$ states have nodes along the $\Gamma$-M direction and, hence, the topological invariant $\nu$ of the 2D system is ill defined. Nonetheless, the nontrivial fictitious 1D invariant $\nu_x$  implies the presence of Majorana modes around $k_x=\pi$ at an interface parallel to the $x$ axis which is confirmed by \figref{Sr2RuO4Spectra}(d). This only holds as long as translation symmetry is preserved along $x$ which might be irrelevant experimentally as very clean samples are already required to stabilize the superconducting state itself.

Taken together, this discussion implies that thin films of Sr$_2$RuO$_4$ represent a promising system for the realization of Majorana modes.

\subsection{Other TRS-breaking condensates}
Let us discuss two further materials, URu$_2$Si$_2$ and UPt$_3$, which are believed to host a TRS-breaking superconducting bulk state \cite{TRSBURu2Si2,TRSBUPt3}.

To begin with URu$_2$Si$_2$, we first note that it is from a symergetic point of view very similar to Sr$_2$RuO$_4$: The point group $D_{4h}$ of its bulk is reduced to $C_{4v}$ in a thin film of (001) orientation and the combination of the strong spin-orbit coupling of U and the small transition temperature \cite{URu2Si2Rev} make the weak-pairing description possible. 
It follows that, if the thin film still displays superconductivity, TRS must be restored in the condensate and the triplet component of the order parameter must be aligned parallel to the plane of the system. In particular, from a pure symmetry point of view, the most natural \cite{SchmalianSubduced} candidate pairing state $e_{(1,i)}$ transforming under the IR $E$ subduced from $E_g$ of $D_{4h}$, which is the IR of the bulk pairing state $\Delta_{\vec{k}} = i\sigma_y(k_x + ik_y)k_z$ \cite{TRSBURu2Si2,ARThermalCond}, is suppressed in the weak-pairing limit. Consequently, superconductivity is likely to disappear in the thin-layer limit.

As for UPt$_3$, the weak-pairing approximation is expected to be applicable for the same reason as in the case of URu$_2$Si$_2$ and Sr$_2$RuO$_4$, but the point symmetries are different: The bulk point group $D_{6h}$ is reduced to $C_{6v}$ in a (001) film and, hence, contains a threefold rotation symmetry such that TRS-breaking cannot be excluded. Due to $C_2^\perp \in C_{6v}$, the triplet vector which is largely aligned along the $z$ axis in the bulk condensate \cite{PauliUPt3} must rotate to be parallel to the $xy$ plane. Note that superconductivity has been reported in (001)-oriented films of UPt$_3$ \cite{UPt3ThinFilms}.
From a symmetry point of view, there are 10 possible superconducting phases -- four associated with the four 1D IRs of $C_{6v}$ and 3 with each of the two 2D IRs $E_1$ and $E_2$. From the symergetic arguments presented above we exclude $E_1$ and two of the 1D IRs as these representations are odd under $C_2^\perp$. Furthermore, the two time-reversal symmetric order parameter configurations transforming under $E_2$ can be discarded and we are left with only 3 candidate pairing states for the thin layer system: The time-reversal symmetric $s$-wave and $i$-wave states transforming under $A_1$ and $A_2$, respectively, as well as the TRS-breaking state transforming as $(X+iY)^2$.

\subsection{Further examples}
Finally, let us discuss two additional 2D superconducting systems without TRS-breaking 3D analogue.
We begin with the LaAlO$_3$/SrTiO$_3$ heterostructures that show interface conductivity \cite{Ohtomo,HerranzDEL,AnnadDEL} for the three different orientations $(001)$, $(110)$, and $(111)$ of the interface with respective point groups $C_{4v}$, $C_{2v}$, and $C_{3v}$ while superconductivity has so far only been reported for the former two orientations \cite{Reyren,110SC}. Due to the strong spin-orbit splitting of the Fermi surfaces \cite{CavigliaSOC,ShalomSOC}, the weak-pairing description is clearly appropriate. 
From our general symergetic arguments, it follows that the condensates of the $(001)$ and $(110)$ interfaces must be necessarily time-reversal symmetric whereas the $(111)$ heterostructure allows for exotic TRS-breaking superconductivity. Due to the absence of a $C_2^\perp$ symmetry it is also the first system we have discussed so far that makes an out-of-plane triplet vector possible. Taken together this motivates a closer experimental inspection of the low-temperature properties of the $(111)$ interface.

In order to calculate the topological invariant $\nu$ in \equref{DefinitionTopInvariant} of the $(001)$ and $(110)$ interfaces, a microscopic calculation has to be performed since there is no 3D analogue to compare with and the symmetry properties alone do not determine $\nu$. The analysis of Refs.~\cite{ScheurerSchmalian,MScheurerMechan} shows that the topological properties are directly related to the origin (electron-phonon/purely electronic) of the interaction driving the superconducting instability. 

Our final example is single-layer FeSe on SrTiO$_3$. 
If the weak-pairing description is also valid for this system, the symergetic restrictions apply and the superconductor cannot transform under the 2D IR of $C_{4v}$, thus, preserving TRS. In combination with experiment indicating the absence of nodes \cite{ARPESFeSe}, there are only three possible pairing states: The pairing field can have the same sign on all four (spin-split) electron pockets around the M point ($s^{++++}$), the signs can be pairwise identical ($s^{++--}$) or only differ on one Fermi surface ($s^{+++-}$). Only the latter pairing state has a nontrivial DIII invariant $\nu$ as readily follows from \equref{DefinitionTopInvariant}. It has recently been shown \cite{MScheurerMechan} under very general assumptions that an $s^{+++-}$ state is not possible irrespective of whether superconductivity arises from the coupling to collective particle-hole modes or from phonons. Therefore, FeSe is most likely a topologically trivial, TRS-preserving superconductor.

\section{Discussion}
The applications of the symergetic selection rules to various materials discussed above show that these can both be used to pinpoint the order parameter of 2D superconductors as well as serve as design principles in the search for superconducting phases with exotic properties such as broken TRS or nontrivial topologies. 
In this context, it is particularly important that our arguments are only based on the assumptions \emph{(i)} and \emph{(ii)} and, hence, go beyond model studies, i.e., do not depend on microscopic details such as number and character of relevant orbitals or the form of the interaction driving the superconducting instability.

Since inversion symmetry is locally broken at the surface of a material, one might wonder whether the symergetic selection rules are also relevant for the behavior of the superconducting phase at the boundary of the system.
In the case of a material such as Sr$_2$RuO$_4$ which consists of weakly coupled layers as illustrated in \figref{DifferentSetups}(c), the condensate near a surface perpendicular to the $z$ axis can be thought of as a set quasi-2D systems with $E_{\text{so}}$ increasing as the distance to the surface is reduced and, hence, bears strong similarities to the superconductor in our gedankenexperiment.
In combination with the near degeneracy \cite{SigristRiceTriplet} with the triplet states transforming under the 1D IRs, we expect the triplet vector to rotate to be parallel to surface and TRS to be restored locally. It is an interesting open question whether this could account for the absence of magnetic signals in experiment \cite{NoSpontCurrents1} that are expected from the chiral $p$-wave nature of the bulk order parameter.

\def\theequation{M\arabic{equation}}
\setcounter{equation}{0}
\begin{footnotesize}
\section{Methods}
 
\subsection{Fourth order of the GL expansion}
Let us provide more details on the proof by contradiction based on the fourth order GL expansion first focusing on real IRs. Expanding \equref{BetaCoefficient} with $\vec{z}^{n_0} \rightarrow \vec{z}^{n_0}_{\xi}= \vec{z}^{n_0}|_{\vec{z}^{n_0}_{\mu_0} \rightarrow \vec{z}^{n_0}_{\mu_0} e^{i\xi}}$ to leading nontrivial order in $\xi$, one finds \equref{AsymptoticForm} with
\begin{equation}
   C(\vec{z}^{n_0}) = 4\sum_{\mu\neq \mu_0} (z^{n_0}_{\mu_0} z^{n_0}_\mu)^2  \sum_a \left\langle \left( \varphi_{\vec{k} a}^{\mu_0 n_0} \varphi_{\vec{k} a}^{\mu n_0} \right)^2 \right\rangle_{\hspace{-0.3em}a\hspace{0.2em}}.  \label{LowerEnergy}
\end{equation} 
To derive \equref{LowerEnergy}, it has been taken into account that $\varphi^{\mu n_0}\in\mathbbm{R}$ and that the symmetries require the free energy to be invariant under 
\begin{align}
  \eta^{n_0}_\mu \rightarrow \begin{cases} - \eta^{n_0}_\mu, & \mu = \mu_0 \\ \eta^{n_0}_\mu, & \mu \neq \mu_0 \end{cases}, \qquad \forall \mu_0 \in \{1,2, \dots ,d_{n_0} \}, \label{reqPropI}
\end{align}
for any real IR $n_0$ of 2D and 3D point groups.
From \equref{LowerEnergy}, we directly see that $C(\vec{z}^{n_0}) > 0$ for all $\vec{z}^{n_0}\neq \vec{e}_{\mu_0}$ unless $\varphi^{\mu n_0} = 0$. In the latter case, however, the superconducting state is fully ungapped in the weak-pairing limit and, hence, disfavored energetically as discussed in the main text.

If $\vec{z}^{n_0} = \vec{e}_{\mu_0}$, we will use $\vec{z}_\xi = (\vec{e}_{\mu_0} + i\xi \vec{e}_{\mu_1} )/\sqrt{1+\xi^2}$ again yielding \equref{AsymptoticForm}, but with modified
\begin{equation}
 C(\vec{z}^{n_0}) = 2\sum_a\left[ \left\langle \left(\varphi_{\vec{k} a}^{\mu_0 n_0}\right)^{4} \right\rangle_{\hspace{-0.3em}a\hspace{0.2em}} - \left\langle \left(\varphi_{\vec{k} a}^{\mu_0 n_0} \varphi_{\vec{k} a}^{\mu_1 n_0}\right)^{2} \right\rangle_{\hspace{-0.3em}a\hspace{0.2em}}\right],
\end{equation} 
which is readily shown to be positive (as long as $\varphi^{\mu n_0}$ are not identically zero).
This completes the proof for the case of real IRs of point groups. 

Due to TRS, complex IRs are always degenerate with their conjugate IR and, hence, can be seen as reducible representations of doubled dimension.
Being reducible, symmetries are less restrictive in this case and, in particular, \equref{reqPropI} is not guaranteed any more which necessitates a generalized form of the proof presented above. The latter can be found in the \supl 3.

\subsection{Inequality for the condensation energy}
Here we discuss how \equref{ConditionForWPL} of the main text is obtained.
To derive a necessary condition for the emergence of a superconducting state with $D_{\vec{k}aa} = 0$, we analyze whether its zero-temperature condensation energy
\begin{equation}
 \Delta E(\Delta_0) = \frac{1}{2} \sum_{a=1}^N\sum_{\vec{k}} \left( \left|E_{\vec{k}a}(\Delta_0)\right| - \left|\epsilon_{\vec{k}a}\right| \right) - \frac{\Delta_0^2}{2g} \label{GeneralExpressionForCondenstionEnergy}
\end{equation} 
is positive for some finite $\Delta_0 = g |\vec{\eta}^{n_0}|$. Here $\epsilon_{\vec{k}a}$ and $E_{\vec{k}a}$ denote the different bands of the normal state and of the superconducting mean-field Hamiltonian, respectively. 

To focus on the essential part of the physics, let us consider only $N=2$ singly degenerate bands. Replacing $|D_{\vec{k}12}|$ by its maximum value $\Delta_0 m$ yields an upper bound $\Delta E^{\text{max}}(\Delta_0)$ on the condensation energy. Physically, it corresponds to the situation of ``optimal basis functions'' with $|D_{\vec{k}12}|$ being constant except for negligibly small regions where it has to vanish as dictated by symmetry.
Evaluating the sum in \equref{GeneralExpressionForCondenstionEnergy} as an integral (cut off energetically at $\Lambda$, constant density of states $\rho_F$) shows that the condensation energy can only be positive when the spin-orbit splitting $E_{\text{so}}$ on the Fermi surface satisfies \equref{ConditionForWPL}, where the dimensionless coupling constant is defined by $\lambda=2\rho_F m^2 g$.
Furthermore, one finds that $\Delta_0 m > E_{\text{so}}/2$ at the positive maximum of $\Delta E^{\text{max}}(\Delta_0)$ revealing the direct connection to assumption \emph{(i)}.

\subsection{Model for Sr$_2$RuO$_4$}
To be self contained, we define the model used in the main text to calculate the spectrum of Sr$_2$RuO$_4$. The centrosymmetric part 
\begin{align}
 h^S_{\vec{k}} = &\begin{pmatrix} \epsilon_{xy}(\vec{k})-\mu-\delta \epsilon_{xy}  & 0 & 0 \\ 0 & \epsilon_{xz}(\vec{k})-\mu & t_\eta \sin(k_1)\sin(k_2) \\ 0 & t_\eta\sin(k_1)\sin(k_2) & \epsilon_{yz}(\vec{k})-\mu  \end{pmatrix} \nonumber \\ &+ \frac{\lambda}{2} \sum_{j=x,y,z} L_j \cdot \sigma_j \label{BulkHamOfSRO} \end{align}
is taken to be of the form usually applied (see, e.g., \refcite{Scaffidi}) to describe the bulk of the material. In \equref{BulkHamOfSRO}, the orbital basis $\{4d_{xy},4d_{xz},4d_{yz}\}$ of Ru orbitals is used and $\sigma_j$ are Pauli matrices representing spin. Furthermore, $\epsilon_{xy}(\vec{k}) = -2t_3\left(\cos(k_1) + \cos(k_2)\right) - 4t_4 \cos(k_1)\cos(k_2)$, $\epsilon_{xz}(\vec{k}) = -2t_1 \cos(k_1) - 2t_2 \cos(k_2)$, and $\epsilon_{yz}(\vec{k}) = -2t_2 \cos(k_1) - 2t_1 \cos(k_2)$. Adding the inversion antisymmetric hopping term $h^A_{\vec{k}} = \delta (L_x \sin(k_y) - L_y \sin(k_x))$ already introduced in the main text defines the normal state Hamiltonian $h^{\phantom{S}}_{\vec{k}} = h^S_{\vec{k}} + h^A_{\vec{k}}$ in \equref{GeneralMFHamiltonian}. To obtain \figref{Sr2RuO4Spectra}(b-d), we have taken $t_2 =0.1t_1$, $t_3 =0.8t_1$, $t_4 =0.3t_1$, $t_\eta = -0.04 t_1$, $\lambda = 0.2t_1$, $\mu=t_1$ and $\delta\epsilon_{xy}=0.1 t_1$ as deduced in \refcite{Scaffidi}.

\end{footnotesize}

\vspace{3em}
\textbf{Acknowledgments:} We thank P.~M.~R.~Brydon and E.~J.~K\"onig for discussions. 

\vspace{1em}





\textbf{Funding:} D.A. is supported by NSF visa grant \mbox{DMREF} 1335215.


%

\onecolumngrid
\def\theequation{S\arabic{equation}}
\setcounter{equation}{0}

\newpage

\section{Supplementary Information on Selection rules for Cooper pairing in two-dimensional interfaces and sheets}

\subsection{S.1 Hermiticity of basis functions}
Here we show that the basis functions $\{\chi_\mu^n\}$ introduced in \equref{eq:expans} to expresses the superconducting order parameter can always be chosen to satisfy the Hermiticity constraint
\begin{equation}
 \left(\chi_{\vec{k}\mu}^n T^\dagger \right)^\dagger = \chi_{\vec{k}\mu}^{\bar{n}}T^\dagger \label{ToBeShown}
\end{equation} 
as a consequence of the TRS of the normal phase. Here $\bar{n}$ denotes the complex conjugate representation of the IR $n$. \equref{ToBeShown} is central for the analysis of the main text as it determines the transformation behavior of the different superconducting states under time-reversal. It also leads to the property $(\varphi_{\vec{k}a}^{\mu n})^* = \varphi_{\vec{k}a}^{\mu \bar{n}}$ of the weak-pairing basis functions $\varphi_{\vec{k}a}^{\mu n}$. The following proof of \equref{ToBeShown} generalizes the results of \refcite{YipGarg} to complex representations and to multiband systems, i.e., goes beyond the pseudospin picture. 

We first generalize the parameterization (\ref{eq:expans}) to
\begin{equation}
 \Delta_{{\vec{k}}} = \sum_n \sum_{\mu=1}^{d_n} \eta_\mu^n \Xi_{\vec{k}\mu}^n(h^n_1,h^n_2) T, \qquad \Xi_{\vec{k}\mu}^n(h_1^n,h_2^n) =  h_1^n \chi_{\vec{k}\mu}^nT^\dagger + h_2^n \left(\chi_{\vec{k}\mu}^{\bar{n}} T^\dagger\right)^\dagger, \qquad \eta_\mu^n, h^n_1, h^n_2 \in \mathbbm{C}. \label{GeneralizedParameterization}
\end{equation} 
It is straightforwardly shown that $\Xi_{\vec{k}\mu}^n(h^n_1,h^n_2)$ transforms exactly as $\chi_{\vec{k}\mu}^{n}T^\dagger$ under $\mathcal{G}$ by taking into account that $[\Theta,R_\psi(g)]=0$ for all $g \in \mathcal{G}$. Upon reparameterization $H^n_1=\frac{1}{2}(h^n_1+h^n_2)$ and $H^n_2=\frac{1}{2i}(h^n_1-h^n_2)$, it holds
\begin{equation}
 \left(\Xi_{\vec{k}\mu}^n(H^n_1,H^n_2)\right)^\dagger = \Xi_{\vec{k}\mu}^{\bar{n}}((H^n_1)^*,(H^n_2)^*). \label{DaggerCondition}
\end{equation} 
Applying \equref{eq:TR} to \equref{GeneralizedParameterization}, one finds that time-reversal is represented by
\begin{equation}
 (\eta_\mu^n,H^n_1,H^n_2) \, \stackrel{\Theta}{\longrightarrow} \, \left(\pm (\eta_\mu^{\bar{n}})^*,(H^{\bar{n}}_1)^*,(H^{\bar{n}}_2)^*\right) \label{TRRep}
\end{equation} 
for $\Theta^2 = \mp \mathbbm{1}$.
The free-energy expansion has again the form (\ref{GLExpansion}) with 
\begin{equation}
 a_n(T;H_1^n,H_2^n) = \sum_{j,j'=1}^2 M^{(n)}_{jj'}(T) (H^n_j)^* H^n_{j'}
\end{equation} 
as follows from gauge invariance and basic scaling arguments. 
Reality of the free energy and invariance under time-reversal (\ref{TRRep}) forces $M^{(n)}$ to be Hermitian and $M^{(\bar{n})} = (M^{(n)})^T$, respectively. It implies that $(\vec{H}^n)^* =\vec{H}^{\bar{n}}$ (without loss of generality) at the minimum of the free energy and, together with \equref{DaggerCondition}, yields \equref{ToBeShown}. In addition, we find from \equref{TRRep} that time-reversal simply becomes
\begin{equation}
  \eta_\mu^n \, \stackrel{\Theta}{\longrightarrow} \, \pm (\eta_\mu^{\bar{n}})^* \label{RepTRSRed}
\end{equation} 
for $\Theta^2 = \mp \mathbbm{1}$ as used in the main text.

\subsection{S.2 Microscopic GL expansion and resummation}
The GL expansion is straightforwardly derived by decoupling the interaction in \equref{InteractinHamInNo} via a Hubbard-Stratonovich transformation in the Cooper channel, integrating out the fermionic degrees of freedom in the weak-pairing limit and subsequent expansion in the superconducting order parameter. One finds
\begin{equation}
 \mathcal{F}[\eta_{\mu}^{n_0}] = \mathcal{F}[0] + \frac{1}{g} \sum_{\mu=1}^{d_{n_0}} |\eta_{\mu}^{n_0}|^2 + \sum_{l=1}^\infty \frac{2^{2l-1}}{l} (-1)^l I_l(\Lambda,T)  \sum_a \left\langle \Bigl|\sum_{\mu =1}^{d_{n_0}} \eta_\mu^{n_0} \varphi_{\vec{k}a}^{\mu n_0}\Bigr|^{2l} \right\rangle_{\hspace{-0.3em}a\hspace{0.2em}} , \label{FinalFormOfGLExp}
\end{equation}
where 
\begin{equation}
   I_l(\Lambda,T) := \rho_F   \int_{-\Lambda}^{\Lambda}\diff\epsilon \, T \sum_{\omega_n} \frac{1}{(\omega^2_n + \epsilon^2)^l} >0 \label{DefinitionOfIntegrals}
\end{equation} 
with $\omega_n$ representing fermionic Matsubara frequencies and $\braket{\dots}_a$ denoting the average over Fermi surface $a$ defined by
\begin{equation}
 \braket{ \dots }_a := \rho_F^{-1} \int_a \diff\Omega_{\vec{k}} \, \rho_a(\Omega_{\vec{k}}) \dots \,. \label{DefinitionOfFSAverage}
\end{equation} 
Here $\rho_F$ and $\rho_a(\Omega)$ are the total and angle-/Fermi-surface-resolved density of states, respectively. Furthermore, $\int_a \diff\Omega_{\vec{k}}$ in \equref{DefinitionOfFSAverage} describes integration over Fermi surface $a$. While the explicit form of $\rho_a(\Omega)$ is irrelevant for the derivation of the selection rules, we only take advantage of positivity $\rho_a(\Omega)>0$.

To study the order of the mean-field transition, we have to focus on the fourth and higher order terms $\mathcal{F}_{\geq 4}$ of the GL expansion which we write as
\begin{equation}
 \mathcal{F}_{\geq 4}[\eta_{\mu}^{n_0}] =  T^2 \sum_a \int_a \diff \Omega_{\vec{k}} \, \rho_a(\Omega_{\vec{k}}) \, f_{\geq 4}\hspace{-0.3em}\left(\Bigl|\sum_{\mu =1}^{d_{n_0}} \eta_\mu^{n_0} \varphi_{\vec{k}a}^{\mu n_0}\Bigr|^{2}/T^2\right).  \label{ResultForFreeEnergy}
\end{equation}
Evaluating $I_l$ in \equref{DefinitionOfIntegrals} (for simplicity in the limit $\Lambda \rightarrow \infty$) leads to a power series representation of the dimensionless function $f_{\geq 4}(x)$ that is positive but only converges for $|x| < \pi^2/4$.

To access larger values of the superconducting order parameter, one can interchange the summation over $l$ and the integration with respect to energy $\epsilon$ in \equref{FinalFormOfGLExp} yielding the analytic continuation on $\mathbbm{R}^+$ of the aforementioned series representation 
\begin{equation}
 f_{\geq 4}(x) = T^{-1} \int_{-\Lambda}^{\Lambda}\diff\epsilon  \sum_{\omega_n} \, g_{\geq 4}\hspace{-0.2em}\left(\frac{x \, T^2}{\omega^2_n + \epsilon^2}\right), \qquad g_{\geq 4}(y)= \frac{1}{2}\left(4y-\ln(1+4y)\right).
\end{equation} 
Due to $g_{\geq 4} > 0$ and, hence, $f_{\geq 4} > 0$ on $\mathbbm{R}^+$, it follows $\mathcal{F}_{\geq 4}[\eta_{\mu}^{n_0}] \geq 0$ as stated in the main text.

\subsection{S.3 Complex IRs}
As mentioned in the main text, complex IRs $n\neq \bar{n}$ deserve further investigation as far as the proof by contradiction based on the fourth order GL expansion is concerned. Due to the representation (\ref{RepTRSRed}) of time-reversal, the TRS of the high-temperature phase implies $a_n = a_{\bar{n}}$ in the GL expansion (\ref{GLExpansion}). Consequently, $n$ and $\bar{n}$ are always degenerate at the quadratic level and are more conveniently treated as a real \textit{reducible} representation $\widetilde{n}$ of dimension $2d_{n}$ with basis functions 
\begin{equation}
 \chi_{\vec{k}\mu}^{\widetilde{n}} = \frac{1}{\sqrt{2}}\left(\chi_{\vec{k}\mu}^{n} + \chi_{\vec{k}\mu}^{\bar{n}}\right), \qquad  \chi_{\vec{k}\mu+d_{n}}^{\widetilde{n}} = \frac{1}{\sqrt{2}\,i}\left(\chi_{\vec{k}\mu}^{n}- \chi_{\vec{k}\mu}^{\bar{n}}\right) \label{DefinitionOfNewChis}
\end{equation}
and associated $\varphi_{\vec{k} a}^{\mu \widetilde{n}} = \braket{\phi_{\vec{k}a}|\chi_{\vec{k}\mu}^{\widetilde{n}}T^{\dagger}|\phi_{\vec{k}a}} \in \mathbbm{R}$.

Using the real representation associated with a pair of complex conjugate representations, one can repeat the proof by contradiction presented in the main text. However, this does not work for all pairs of complex IRs of the point groups of crystalline 2D and 3D systems:
For the complex IRs of $C_4$, $S_4$ and $C_{4h}$, all of which are 1D, the associated 2D real representation does not ensure invariance under \equref{reqPropI}.
Here, the symmetries only impose the constraints $\sum_a\braket{(\varphi^{\mu\widetilde{n}}_{\vec{k}a})^4}_a\equiv\alpha$ independent of $\mu=1,2$ and $\gamma_{12} = -\gamma_{21} \equiv \gamma$ with $\gamma_{\mu\mu'}=\sum_a\braket{(\varphi^{\mu\widetilde{n}}_{\vec{k}a})^3\varphi^{\mu'\widetilde{n}}_{\vec{k}a}}_a$. Inserting $\vec{z}_\xi^{\widetilde{n}} = (\cos(\phi)e^{i\xi},\sin(\phi))$ in \equref{BetaCoefficient} and expanding in $\xi$ yields \equref{AsymptoticForm} with $C=\beta f_1(\phi,\gamma/\beta)$ where
\begin{equation}
 f_1(\phi,g_1) = \sin(2\phi) \left(\sin(2\phi)+ g_1 \cos(2\phi)\right)
\end{equation} 
and $\beta = \sum_a\braket{(\varphi^{1\widetilde{n}}_{\vec{k}a}\varphi^{2\widetilde{n}}_{\vec{k}a})^2}_a$. For any nonzero $g_1$, one can find $\phi\in\mathbbm{R}$ with $f_1(\phi,g_1) < 0$ indicating that the proof by contradiction of the main text has to be extended in order to exclude real-valued order parameter vectors. 

To this end, we consider a generalized transformation $\vec{z}_\xi^{\widetilde{n}} \rightarrow e^{i \vec{\xi}\cdot \vec{\tau}}\vec{z}_\xi^{\widetilde{n}}$ with $\vec{\tau}=(\tau_x,\tau_y,\tau_z)$ denoting Pauli matrices. So far, we have focused on $\vec{\xi} = \xi\vec{e}_3$. Choosing $\vec{\xi} = \xi\vec{e}_1$ instead, we obtain $C= \frac{\alpha-\beta}{2}f_2(\phi,2\gamma/(\alpha-\beta))$ where
\begin{equation}
 f_2(\phi,g_2) = \cos(2\phi) \left(\cos(2\phi)+ g_2 \sin(2\phi)\right).
\end{equation} 
One can show there is no $\phi\in\mathbbm{R}$ with both $f_1(\phi,g_1)$, $f_2(\phi,g_2)$ being non-positive unless
\begin{equation}
 g_1g_1 \geq 1 \quad\Leftrightarrow \quad \gamma^2 \geq \frac{1}{2} \beta(\alpha-\beta). \label{ConditionForProofFail}
\end{equation}  
\equref{ConditionForProofFail} can be shown to be not satisfied by noting that $\braket{\varphi^{\mu}|\varphi^{\mu'}} := \sum_a\braket{ \varphi^{\mu}_{\vec{k}a} \varphi^{\mu'}_{\vec{k}a} }_a$ defines an inner product and, hence, obeys the Cauchy-Schwarz inequality. The latter leads to
\begin{equation}
  \gamma^2 = \frac{1}{4} \left|\braket{\varphi^{1\widetilde{n}}\varphi^{2\widetilde{n}}|(\varphi^{1\widetilde{n}})^2-(\varphi^{2\widetilde{n}})^2}\right|^2 \leq \frac{1}{4} \braket{\varphi^{1\widetilde{n}}\varphi^{2\widetilde{n}}|\varphi^{1\widetilde{n}}\varphi^{2\widetilde{n}}}\cdot\braket{(\varphi^{1\widetilde{n}})^2-(\varphi^{2\widetilde{n}})^2|(\varphi^{1\widetilde{n}})^2-(\varphi^{2\widetilde{n}})^2} = \frac{1}{2} \beta (\alpha-\beta).
\end{equation} 
Consequently, in the case of a complex IR, the superconducting order parameter must necessarily break TRS if the weak-pairing description applies.

\subsection{S.4 Orientation of the triplet vector}
Finally, we present more details on the symergetic restrictions on the triplet vector in the case of several orbitals and/or subbands. In general, the decomposition into singlet and triplet assumes the form
\begin{equation}
 \Delta_{\vec{k}} = \Delta^S_{\vec{k}} \, T + \sum_{j=x,y,z} D^j_{\vec{k}} \, \sigma_j T \label{ParameterizationOfSingletAndTriplet}
\end{equation} 
with both $\Delta^S_{\vec{k}}$ and $D^j_{\vec{k}}$ being matrices in orbital/subband space. For concreteness, we choose a real orbital basis for which $T = \mathbbm{1} \otimes i\sigma_y$. Fermi statistics implies
\begin{equation}
  \left(\Delta^S_{-\vec{k}}\right)^T = \Delta^S_{\vec{k}} \qquad  \left(D^j_{-\vec{k}}\right)^T  = -D^j_{\vec{k}}. \label{FermiDiracConstraintsSimplf}
\end{equation} 
Focusing on the relevant situation $C_2^\perp\in\mathcal{G}$, the order parameter must be even under $C_2^\perp$ and, hence, satisfy
\begin{equation}
  R_{\psi_o}\hspace{-0.1em}(C_2^\perp) \Delta^S_{-\vec{k}} R^\dagger_{\psi_o}\hspace{-0.1em}(C_2^\perp) = \Delta^S_{\vec{k}}, \quad R_{\psi_o}\hspace{-0.1em}(C_2^\perp) D^{1,2}_{-\vec{k}} R^\dagger_{\psi_o}\hspace{-0.1em}(C_2^\perp) = -D^{1,2}_{\vec{k}}, \quad  R_{\psi_o}\hspace{-0.1em}(C_2^\perp) D^{3}_{-\vec{k}} R^\dagger_{\psi_o}\hspace{-0.1em}(C_2^\perp) = D^{3}_{\vec{k}} \label{ConditionsOfRotation}
\end{equation} 
with $R_{\psi_o}\hspace{-0.1em}(C_2^\perp)$ denoting the representation of $C_2^\perp$ in orbital space. 

To proceed further, we take advantage of the fact that the eigenspaces of the spin-independent part of the unit cell Hamiltonian are spanned by basis functions of the IRs of $\mathcal{G}$ (including spin-orbit coupling would require considering the IRs of the associated double group). For example, in case of Sr$_2$RuO$_4$ the relevant low-energy subspaces are spanned by the $4d_{xy}$ ($B_2$ of $C_{4v}$) and $4d_{xz}$, $4d_{yz}$ ($E$ of $C_{4v}$) orbitals. This basis is most convenient for discussing the consequences for the triplet component of the order parameter as $R_{\psi_o}\hspace{-0.1em}(C_2^\perp)$ simply becomes
\begin{equation}
 R_{\psi_o}\hspace{-0.1em}(C_2^\perp) = \text{diag}\left(c_1,c_2,\dots ,c_{N/2}\right), \qquad c_j \in \{+1,-1\},
\end{equation} 
in this basis. Using this in \equref{ConditionsOfRotation} in combination with the Fermi constraint (\ref{FermiDiracConstraintsSimplf}), it directly follows that all diagonal components of $D^3_{\vec{k}}$ vanish identically as stated in the main text.

\end{document}